\documentclass{article}

\usepackage{arxiv}

\usepackage[utf8]{inputenc} 
\usepackage[T1]{fontenc}    
\usepackage{hyperref}       
\usepackage{url}            
\usepackage{booktabs}       
\usepackage{amsfonts}       
\usepackage{nicefrac}       
\usepackage{microtype}      
\usepackage{lipsum}
\usepackage{amsmath,amssymb,amsfonts,amsbsy}
\usepackage{algorithmic}
\usepackage{graphicx}
\usepackage{xcolor}
\usepackage{textcomp}
\usepackage{subfigure}
\usepackage{adjustbox}
\usepackage{comment}
\usepackage[]{algorithm2e}
\usepackage{float}
\usepackage{multirow}
\usepackage{mathrsfs} 
\usepackage{doi}

\newcommand{\bs}[1]{\boldsymbol{#1}}

\DeclareMathOperator*{\argmax}{arg\,max}
\title{Study on Spike-and-wave detection in epileptic signals using t-location-scale distribution and the $k$-nearest neighbors classifier}

\author{
  Antonio~Quintero-Rinc\'on, Valeria Muro, Carlos D'Giano\\
  Epilepsy and Telemetry Integral Center\\
  Foundation for the Fight against Pediatric Neurological Disease (FLENI), Argentina \\
   \And
Jorge~Prendes \\
  University of Toulouse, IRIT-INP-ENSEEIHT, Toulouse, France\\
  \And
  To cite this work, please use this reference
  \doi{10.1109/URUCON.2017.8171869}
}

\begin{document}
\maketitle

\begin{abstract}
Pattern classification in electroencephalography (EEG) signals is an important problem in biomedical engineering since it enables the detection of brain activity, particularly the early detection of epileptic seizures. In this paper, we propose a $k$-nearest neighbors classification for epileptic EEG signals based on a t-location-scale statistical representation to detect spike-and-waves. The proposed approach is demonstrated on a real dataset containing both spike-and-wave events and normal brain function signals, where our performance is evaluated in terms of classification accuracy, sensitivity, and specificity. 
\end{abstract}


\section{Introduction}
\label{sec:intro}
Epilepsy is a chronic disorder resulting from the hyperexcitability of neurons. The electroencephalogram (EEG) is the premier diagnostic tool for epilepsy and provides a key element for the classification and detection of epileptic seizures. The information about the morphology and dynamics of EEG signals can be used to accurately identify seizure onset and quantify the severity and dynamical progression of seizure activity.

The most relevant EEG characteristics employed to classify epileptogenic abnormality can be categorized in terms of spectral properties, signal morphology, and statistical measures \cite{EpilepsyIntersection2011}. In this work, we will focus on signal morphology, specifically in spike-and-wave discharge (SWD) \cite{Blume1988}, which is an EEG generalized discharge pattern seen particularly during absence epilepsy,  whose clinical importance lies in cognitive and behavioral disturbances \cite{AtlasEpilepsy2007}, and its waveform has a regular and symmetric morphology as seen in the Fig.\ref{fig:swd}.

The t-location-scale distribution or non-standardized Student's t-distribution is a statistical model for univariate and multivariate signals that describe its features through three parameters estimated by maximum likelihood: location, shape, and a non-negative scale. It is useful for modeling data distributions with heavier tails that are more prone to outliers than the normal distribution. The t location-scale distribution has been applied to different signal processing problems in diverse areas such as radar, watermark, speech, and wireless; in medicine and health it is widely used in genetics and recently used in sleep patterns \cite{Ong2016}.

The $k$-nearest neighbor (kNN) algorithm is a set of local models, each defined using a single instance, the goal is to return the majority target level within the set of $k$-nearest neighbors to the query, to dilute the dependency of the algorithm on individual instances, such as noise \cite{FundMLearning2015}. kNN is useful in many machine learning and data mining areas, such as bioinformatics, image processing, data compression, document retrieval, computer vision, multimedia databases, and marketing data analysis. See   \cite{Acharya2013,Siuly2015,Zhang2015} for some works in epilepsy and EEG signals.

The rest of this paper is organized as follows: Section \ref{sec:meth} describes the proposed methodology that combines a t-location-scale statistical model and a $k$-nearest neighbors classification scheme for epileptic EEG data. In Section \ref{sec:res} the proposed methodology is applied to real EEG signals from patients suffering from epilepsy to detect spike-and-wave patterns. Conclusions and perspectives for future work are finally reported in Section \ref{sec:disc}.

\section{Methodology}
\label{sec:meth}
Let $\bs X \in \mathbb{R}^{N\times M}$ denote the matrix gathering $M$ EEG signals measured simultaneously on different channels and at $N$ discrete time instants. 

The proposed methodology is composed of three stages. The first stage splits the original signal $\bs X$ into a set of non-overlapping segments using a rectangular sliding window $\bs \Omega = \bs \Omega_0 \left( w-\frac{W-1}{2}\right)$ with $0 \leq w\leq W-1$, so that $\bs X_i = \bs \Omega_{i} \bs X$. In the second stage, the parameters of the t-location-scale distribution for each $\bs X_i$ are estimated. Finally, in the third stage, the feature vector associated with each time segment is classified by using a  $k$-nearest neighbors classifier as \emph{spike-and-wave/non-spike-and-wave}. 

We now introduce the t-location-scale statistical model and the $k$-nearest neighbors classifier used in this paper.

\subsection{t-location-scale distribution}
\label{ssec:dist}
The t-location-scale distribution is a statistical model that belongs to a location-scale family, formed by translation and rescaling of the Student's t-distribution.

The probability density function (PDF) of a location-scale distribution, is given by

\begin{align}
	g(x|\mu,\sigma) = \frac{1}{\sigma} \psi \left(\frac{x-\mu}{\sigma}\right)
\label{eq:ls}	
\end{align}

The probability density function (PDF) of the Student's t-distribution, is given by
\begin{align}
\psi=\frac{\Gamma\left(\frac{\nu+1}{2}\right)}{\sqrt{\nu\pi}\; \Gamma \left(\frac{\nu}{2}\right)} \left[\frac{\nu+x^2}{\nu}\right]^{-\left(\frac{\nu+1}{2}\right)}
\label{eq:st}
\end{align}
Therefore applying \eqref{eq:st} to \eqref{eq:ls}, we have the probability density function (PDF) of the t-location-scale PDF, which is given by
\begin{align}
\label{eq:PDF}
f_\textnormal{tls}(x |\mu,\sigma,\nu) = \frac{\Gamma\left(\frac{\nu+1}{2}\right)}{\sigma\sqrt{\nu\pi}\; \Gamma \left(\frac{\nu}{2}\right)} \left[\frac{\nu+\left(\frac{x-\mu}{\sigma}\right)^2}{\nu}\right]^{-\left(\frac{\nu+1}{2}\right)}
\end{align}
where $-\infty < \mu < \infty$ is the location parameter, $\sigma>0$ is the  scale parameter, $\nu>0$  is the shape parameter, and $\Gamma(.)$ is the Gamma function.

The optimization problem can then be solved using the simplex search method of Lagarias et al. \cite{Lagarias1998}. We refer the reader to \cite{Hazraa2017} for a comprehensive treatment of the mathematical properties of the scale-location distribution.

\subsection{$k$-nearest neighbors Classifier (kNN)}
\label{ssec:knnclass}

The $k$-nearest neighbor algorithm is a set of local models, each defined using a single instance. Given a training dataset consisting of vectors $\bs X_i$ and classes $\bs T_i$, kNN returns the majority target level within the set of $k$-nearest neighbors to the query $\bs q$, so that
\begin{align}
	\mathbb{M}_k(\bs q) = \argmax_{\ell \;\in \; \textrm{class}(t)} \sum_{i=1}^{k} \delta(t_i,\ell)
\end{align}

where $\mathbb{M}_k(\bs q)$ is the prediction of the model $\mathbb{M}$ for the query $\bs q$ given the parameter of the model $k$, $\textrm{class}(t)$ is the set of two possible classes in the domain of the target feature, spike-and-wave and non-spike-and-wave and $\ell$ is an element of this set; $i$ iterates over the instances $\bs X_i$ in increasing distance from the query $\bs q$; $t_i$ is the value of the target feature for instance (we would like to remind the reader that $\bs X_i$ denotes each segment of the evaluated signal); and $\delta(t_i, \ell)$ is the Kronecker delta function, which takes two parameters and returns $1$ if they are equal and $0$ otherwise regularizing the decision boundary for the dataset. 
We refer the reader to \cite{Bishop2006,Muja2009,FundMLearning2015} for a comprehensive treatment
of the mathematical properties of $k$-nearest neighbors.

\section{Results}
\label{sec:res}
\begin{figure}[H]
\centering
\includegraphics[width=80mm]{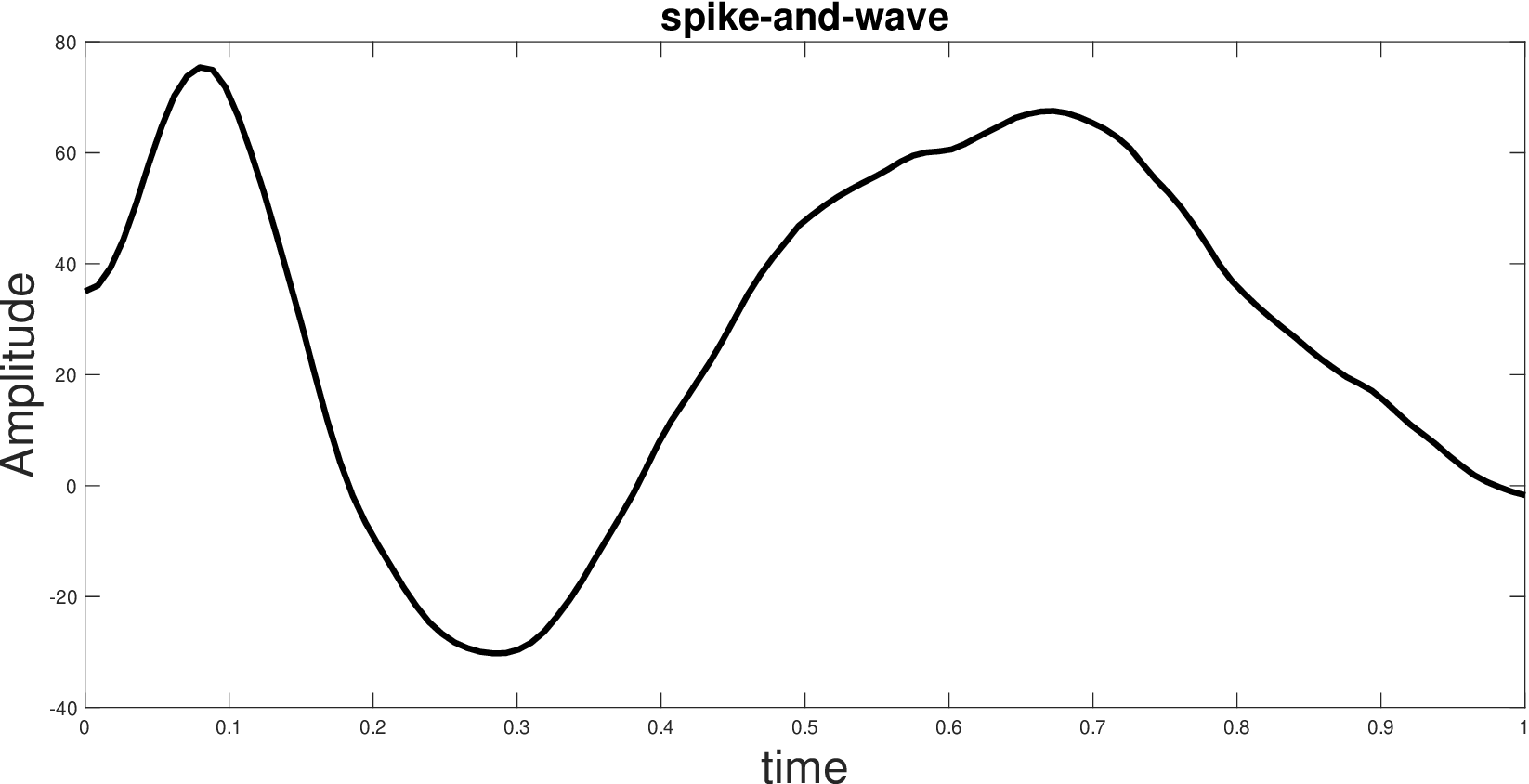}
\caption{spike-and-wave example, we can see its symmetric and regular morphology.}
\label{fig:swd}
\end{figure}

\begin{figure}[H]
\centering
\subfigure{\includegraphics[width=100mm]{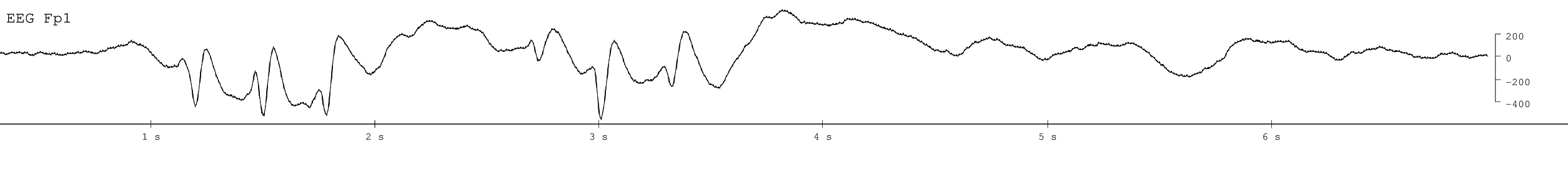}}
\subfigure{\includegraphics[width=100mm]{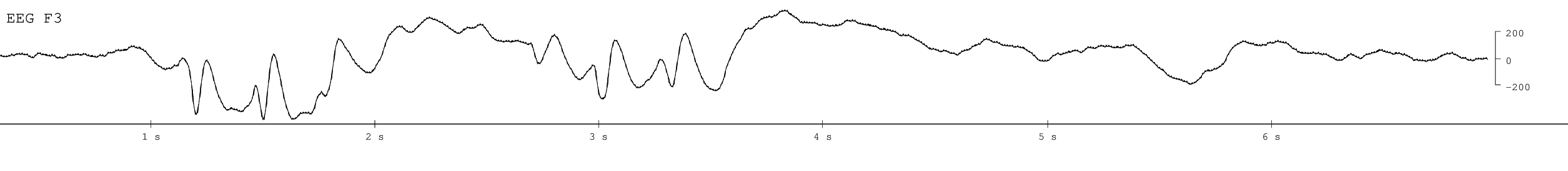}}
\subfigure{\includegraphics[width=100mm]{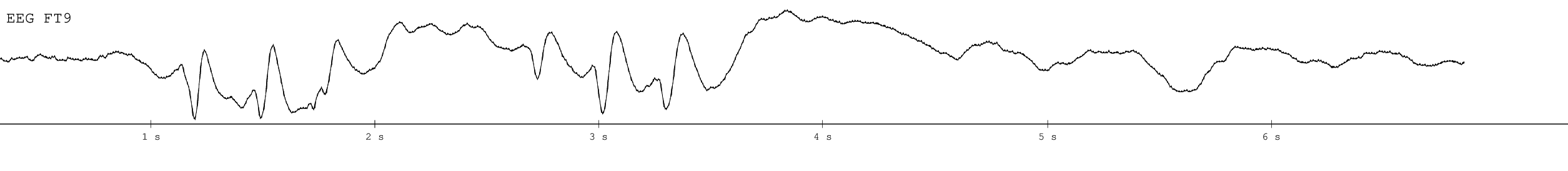}}
\caption{Spikes-and-waves examples in clinical environment. Amplitude (y-axis) in mV and time (x-axis) in sec.}
\label{fig:epochs}
\end{figure}

In this preliminary work, we created a database with 192 monopolar 256Hz signals for an offline training classifier, 96 spike-and-waves and 96 non-spikes-and-waves, measured from two patients. Note that, the spike-and-waves have different times and waveforms, but their morphology is preserved, while the non-spike-and-waves have normal waveforms, see Fig.\ref{fig:swd} and Fig. \ref{fig:epochs}. The goal is to train and subsequently test the capacity of our kNN classification scheme to identify spike-and-wave and non-spike-and-wave in EEG long-time signals. To assess the performance of the proposed method, we used a set of 46 new test monopolar signals; which correspond to each channel of 2 epochs from EEG long-time recordings from one subject. Each monopolar signal uses the following channels: Fp1, Fp2, F7, F3, Fz, F4, F8, T3, C3, Cz, C4, T4, T5, P3, Pz, P4, T6, O1, O2, Oz, FT10, and FT9. 
Each recording contains different spike-and-waves events. Their onset and duration time have been labeled by an expert neurologist. Here we used the expert annotations to extract a short epoch from each recording such that it is focused on the spike-and-wave in long-time signals (the epochs have a duration of the order of 1 minute). 

\begin{figure}[H]
\centering
\subfigure[Location ($\mu$) vs. scale ($\sigma$)]{\hspace*{-0.6cm}\includegraphics[width=70mm]{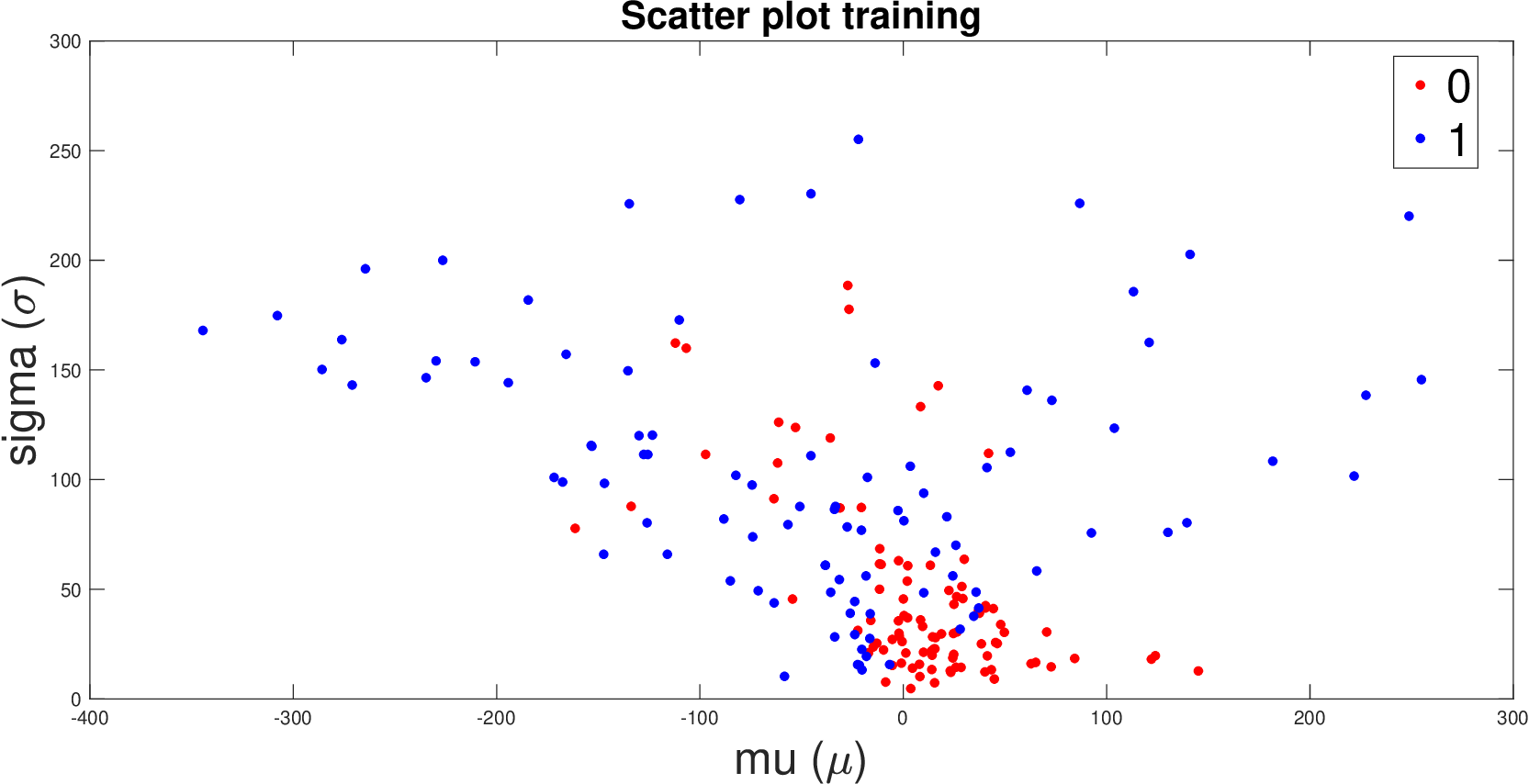}}
\subfigure[Location ($\mu$) vs. shape ($\nu$)]{\includegraphics[width=70mm]{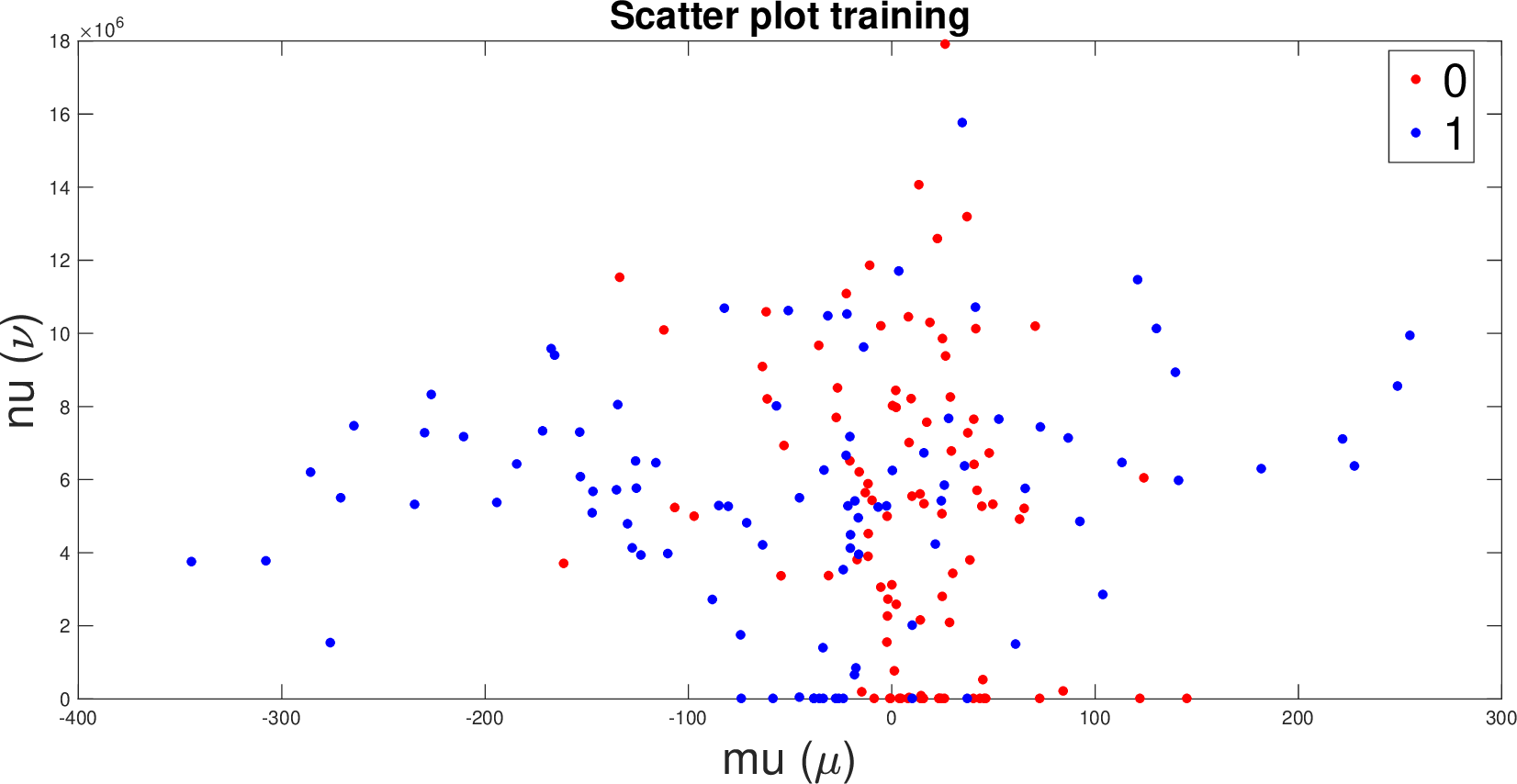}}
\subfigure[Scale ($\sigma$) vs. shape ($\nu$)]{\includegraphics[width=70mm]{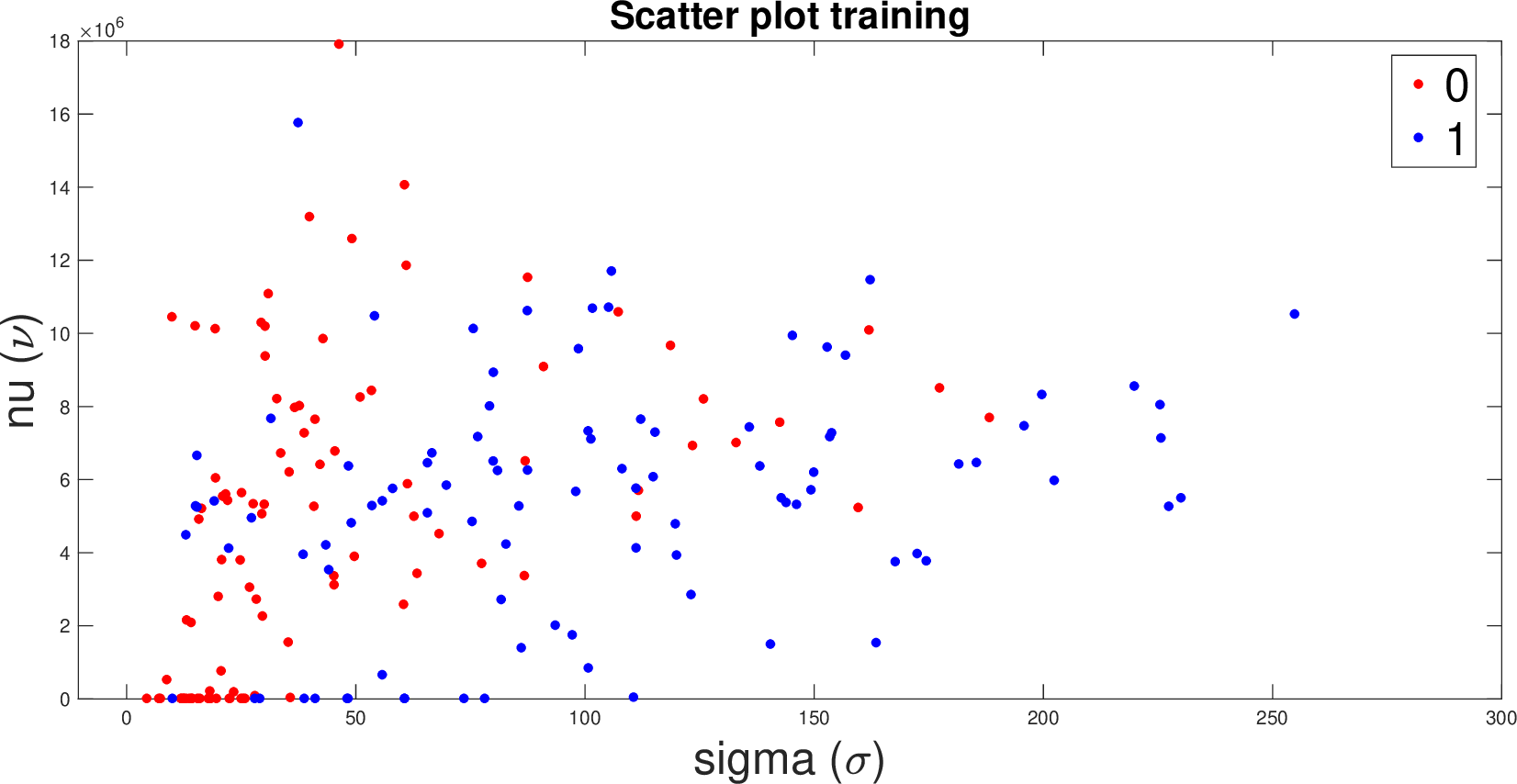}}
\caption{Scatter plots of the offline training classification in 192 dataset signals, for the t-location-scale parameters $\mu$, $\sigma$ and $\nu$ for spike-and-waves events (blue dots) and non-spike-and-waves events (red dots), showing the data dispersion of the proposed approach. In a) and c) spike-and-waves tend to have a higher scale $\sigma$, in b) non-spikes-and-waves tend to have a location $\mu$ concentrated between 0 and 100.}
\label{fig:scatterT1}
\end{figure}

\begin{figure}[H]
\centering
\subfigure[Location ($\mu$) vs. scale ($\sigma$)]{\hspace*{-0.7cm}\includegraphics[width=70mm]{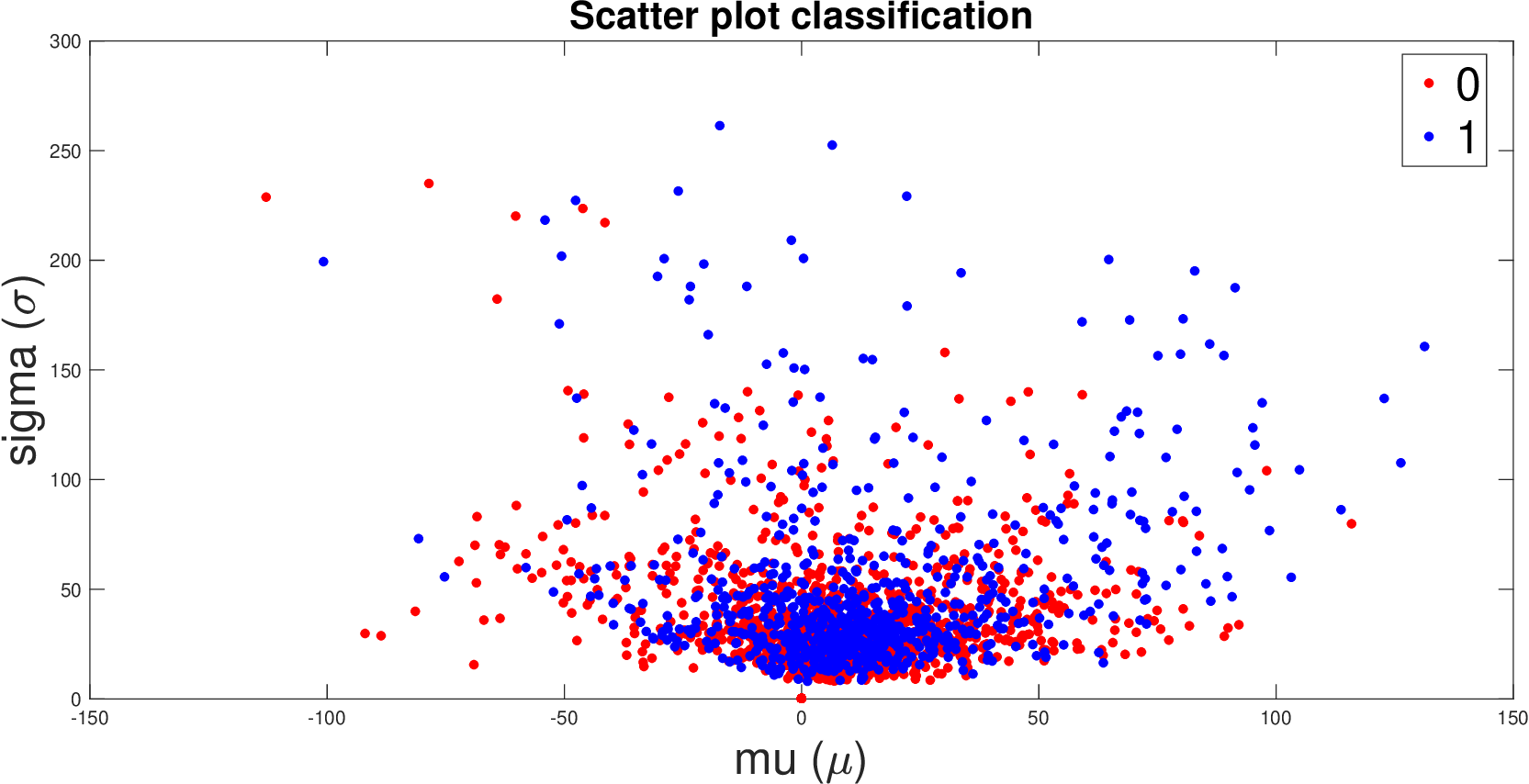}}
\subfigure[Location ($\mu$) vs. shape ($\nu$)]{\includegraphics[width=70mm]{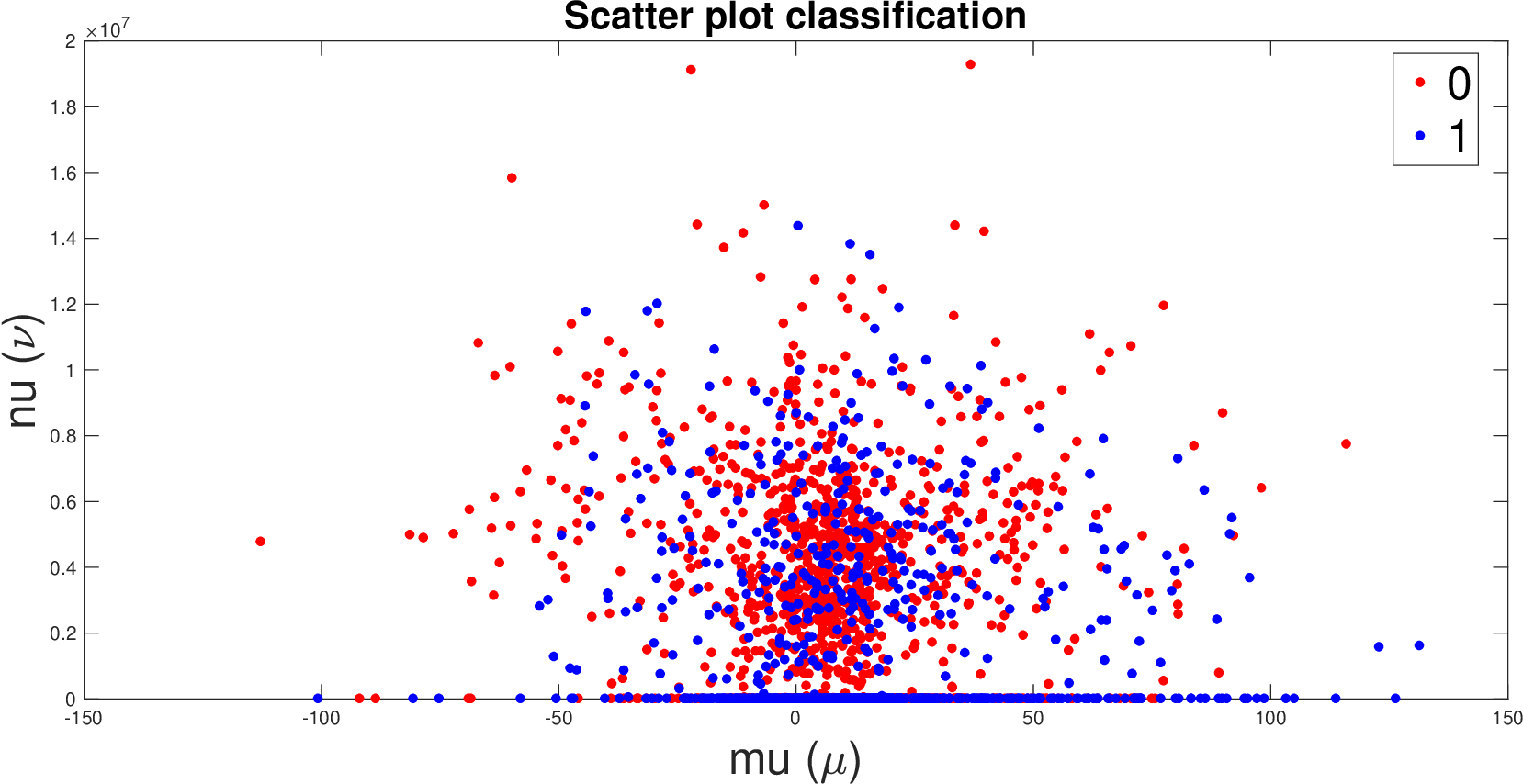}}
\subfigure[Scale ($\sigma$) vs. shape ($\nu$)]{\includegraphics[width=70mm]{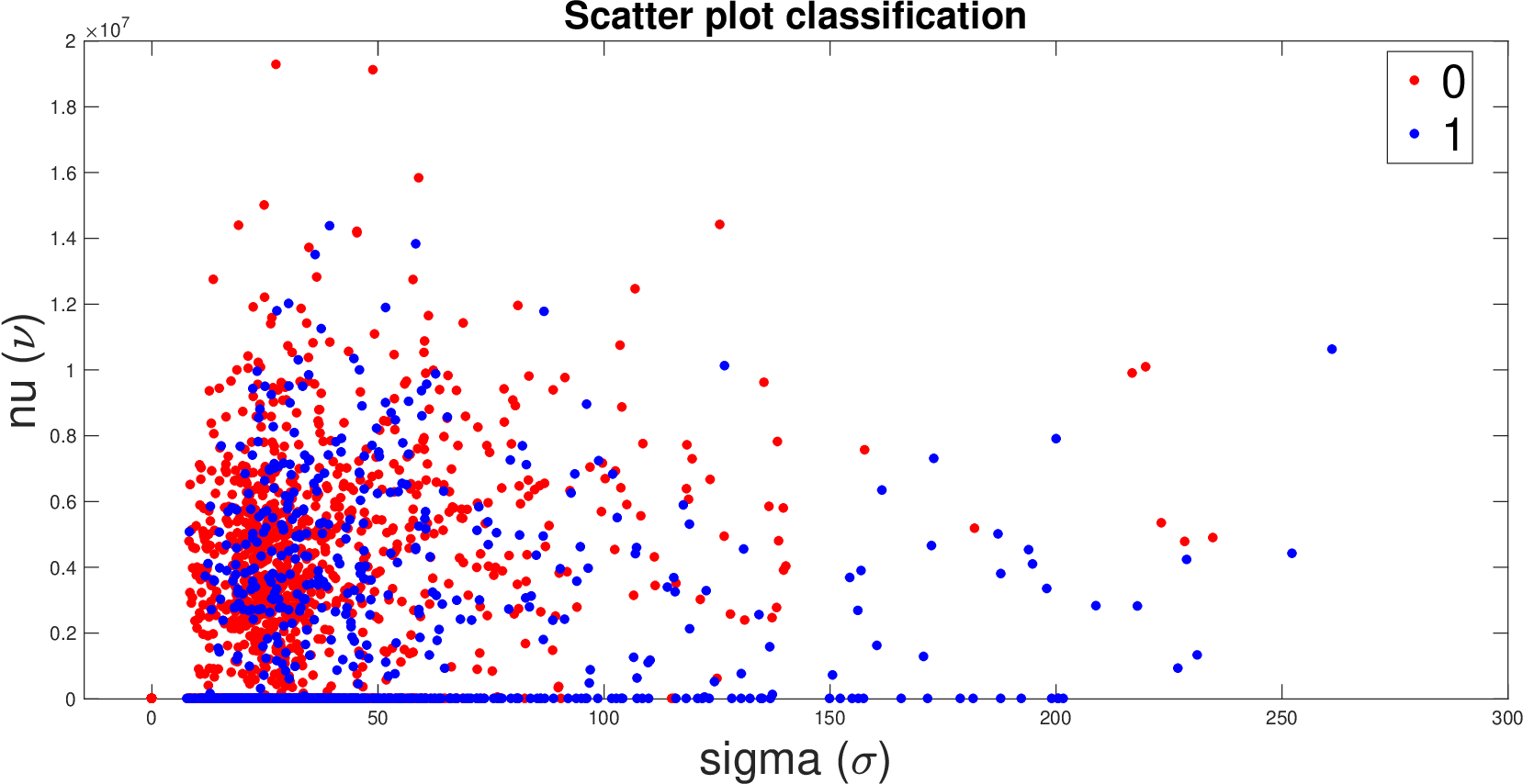}}
\caption{Scatter plots in the online classification of 46 test signals, for the t-location-scale parameters $\mu$, $\sigma$ and $\nu$ for spike-and-waves events (blue dots) and non-spike-and-waves events (red dots), showing the data dispersion of the proposed approach. In a) spike-and-waves tend towards the center down, in b) the trend is not very clear, although there is a great concentration of spike-and-waves in the center down near zero, and c) spike-and-waves tend to be located towards the right and near zero.}
\label{fig:scatterC1}
\end{figure}

Fig. \ref{fig:scatterT1} and Fig. \ref{fig:scatterC1} show scatter plots of the kNN offline training classifier and online classification respectively, whose parameters were estimated using the three t-location-scale distribution parameters: location ($\mu$), scale ($\sigma$) and shape ($\nu$). Data dispersion of spike-and-waves events (label 1: blue dots) and non-spike-and-waves events (label 0: red dots), in Fig. \ref{fig:scatterT1} during the training stage, we can suggest that in a) and c) spike-and-waves tend to have a higher scale $\sigma$ for non-spike-and-waves, in b) non-spikes-and-waves tend to have a location $\mu$ concentrated between a certain threshold for spike-and-waves; while the data dispersion in Fig. \ref{fig:scatterC1}  during the validation stage, we can suggest that in a) spike-and-waves tend towards the center down for non-spike-and-waves, in b) the trend is not very clear, although there is a great concentration of spike-and-waves in the center down near zero for non-spike-and-waves, and in c) spike-and-waves tend to be located towards the right and near zero for non-spike-and-waves.

\begin{figure}[H]
\centering
\includegraphics[width=180mm]{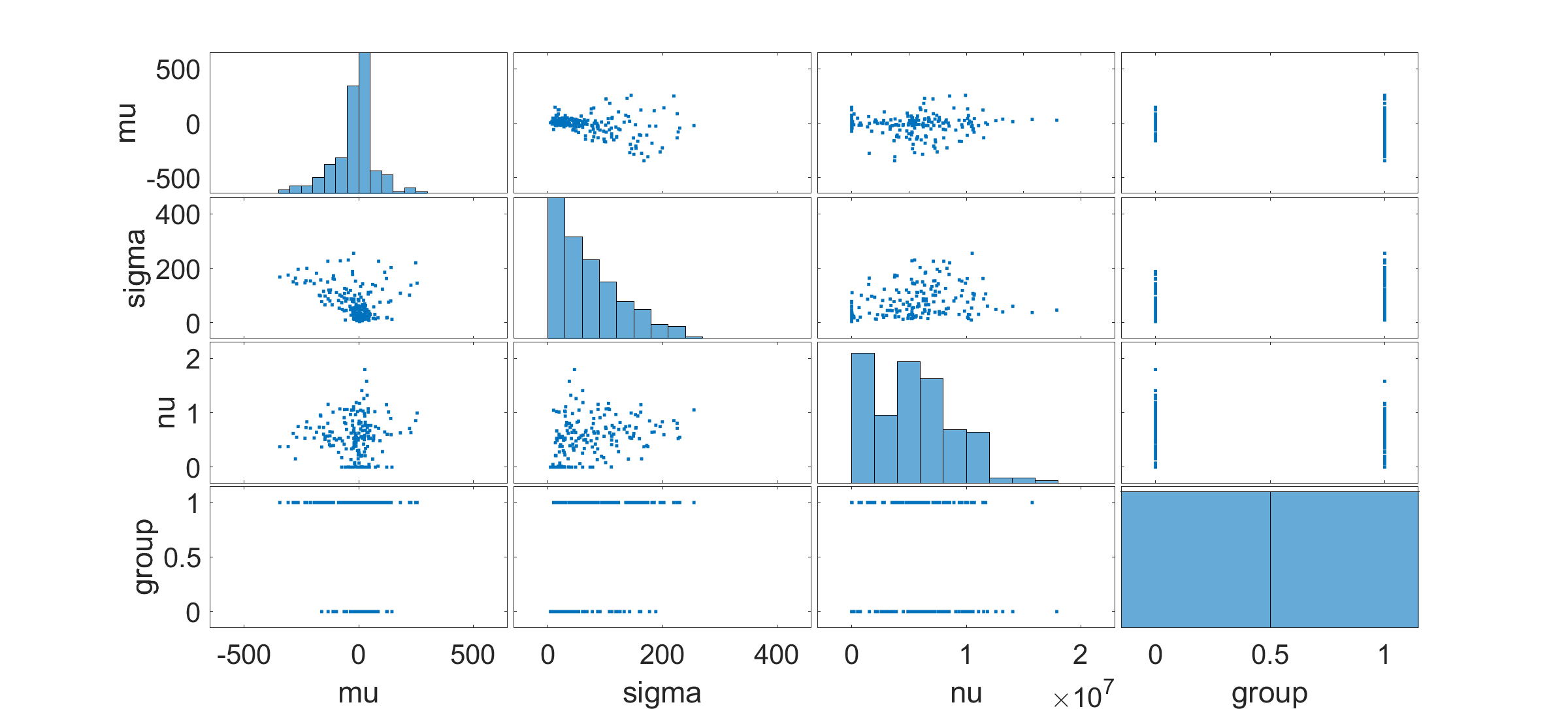}
\caption{Scatter plot in offline training classification in 192 dataset signals for the t-location-scale parameters $\mu$, $\sigma$, and $\nu$, we can see the discrimination between two groups whose size is the same (96 spike-and-waves and 96 non-spikes-and-waves), label 1 for spike-and-wave and label 0 for non-spike-and-wave.}
\label{fig:scatterT2}
\end{figure}

\begin{figure}[H]
\centering
\includegraphics[width=180mm]{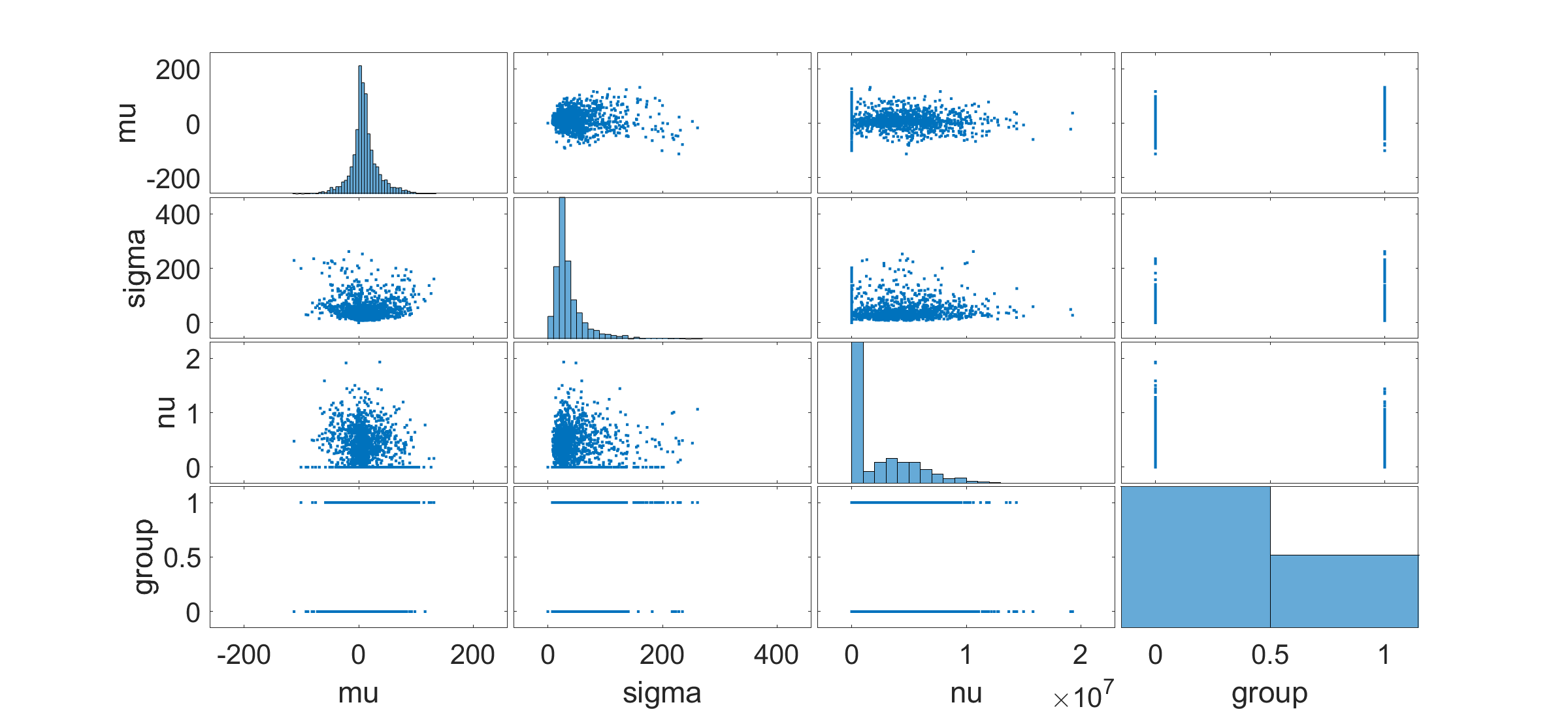}
\caption{Scatter plot in online classification in 46 test signals for the t-location-scale parameters $\mu$, $\sigma$, and $\nu$, we can see the discrimination between two groups whose size is different (spike-and-waves labeled by an expert neurologist and non-spikes-and-waves), label 1 for spike-and-wave and label 0 for non-spike-and-wave.}
\label{fig:scatterC2}
\end{figure}

From the illustration in Fig. \ref{fig:scatterT2} and Fig. \ref{fig:scatterC2} we can see the different histograms and the group discrimination between spike-and-waves events (label 1) and non-spike-and-waves events (label 0) in offline training classification for Fig. \ref{fig:scatterT2} and online classification Fig. \ref{fig:scatterC2}

The performance of the online $k$-nearest neighbors classification method using an equal weight distance with number of neighbors equal to one and distance metric Euclidean, see Fig. \ref{fig:scatterC1} and Fig. \ref{fig:scatterC2}, was assessed in terms of overall accuracy classification, and achieves a 100\% of sensitivity (True Positive Rate) and specificity (True Negative Rate) for spike-and-wave detection.

\section{Conclusion}
\label{sec:disc}
This work presented a new classification method to detect spikes-and-waves events in EEG signals. The method is based on the t-location-scale statistical model and a  $k$-nearest neighbors classification scheme that discriminates spike-and-wave from non-spike-and-wave events. The performance of the proposed method was evaluated on a real dataset containing 192 signal recordings related to both spike-and-wave and non-spike-and-wave events used for offline training classification and 46 labeled test signals used for online classification, achieving a 100\% sensitivity (True Positive Rate) and specificity (True Negative Rate) for spike-and-wave detection. Future work will focus on an extensive evaluation of the proposed approach, comparison with other distributions, automatic count of spike-and-waves in the online classification of EEG long-term signals during sleep, and increase the database with new spike-and-waves found in new test signals.


\end{document}